\documentclass[usenatbib,a4paper,times,fleqn]{mnras}
\usepackage{graphicx,natbib,times,amssymb,amsmath,pstricks}
\usepackage{aas_macros}
\usepackage{bm,url}
\usepackage{textcomp}
\usepackage{breakurl}
\usepackage{subfig}

\title[Horseshoes in transitional discs]{On the origin of horseshoes in transitional discs}

\author[Ragusa et al.]{{Enrico Ragusa$^{1}$\thanks{enrico.ragusa@unimi.it}, 
Giovanni Dipierro$^{1}$\thanks{giovanni.dipierro@unimi.it},
Giuseppe Lodato$^{1}$,
Guillaume Laibe$^{2}$ \&
Daniel J. Price$^{3}$} \\
$^{1}$Dipartimento di Fisica, Universit\`a Degli Studi di Milano, Via Celoria, 16, Milano, I-20133, Italy \\
$^{2}$School of Physics and Astronomy, University of St. Andrews, North Haugh, St. Andrews, Fife KY16 9SS, UK \\
$^{3}$Monash Centre for Astrophysics (MoCA) and School of Physics and Astronomy, Monash University, Clayton Vic 3800, Australia
}
\pagerange{\pageref{firstpage}--\pageref{lastpage}} \pubyear{2016}

\date{}
\begin{document}
\label{firstpage}
\bibliographystyle{mnras}
\maketitle

\begin{abstract}
We investigate whether the rings, lopsided features and horseshoes observed at millimetre wavelengths in transitional discs can be explained by the dynamics of gas and dust at the edge of the cavity in circumbinary discs. We use 3D dusty smoothed particle hydrodynamics calculations to show that binaries with mass ratio $q \gtrsim 0.04$ drive eccentricity in the central cavity, naturally leading to a crescent-like feature in the gas density, which is accentuated in the mm dust grain population with intensity contrasts in mm-continuum emission of 10 or higher. We perform mock observations to demonstrate that these features closely match those observed by ALMA, suggesting that the origin of rings, dust horseshoes and other non-axisymmetric structures in transition discs can be explained by the presence of massive companions. 
\end{abstract}


\section{Introduction}
Recent spectacular observations of dust and gas in nearby protoplanetary discs have revealed substructures in the form of spirals, gaps, cavities and ring-like features (\citealt{alma-partnership15a,andrews16a}; see recent review by \citealt{casassus16a}). Whether, and how, such structures are created is critical to understanding the planet formation process.

 One of the most spectacular first results with the Atacama Large Millimetre/Submillimetre Array (ALMA) was the observation of a non-axisymmetric `horseshoe' in the dust continuum emission in Oph IRS 48 \citep{vandermarel13}, with subsequent observations revealing asymmetric structures in several other transition discs \citealt{vandermarel16a}. Such features are most commonly interpreted (including by \citealt{vandermarel13}) as vortices, for example arising from the Rossby Wave Instability (RWI) at the edge of the gap formed by a young planet \citep{lovelace99,lyra09,lyra13a,zhu14b} or as a result of internal dynamical processes associated with the presence of a weak magnetic field \citep{ruge16a}. RWI arises in sufficiently inviscid discs, with equivalent $\alpha$ parameters \citep{shakura73} of the order of $\alpha\lesssim 10^{-4}$. Such a low viscosity allows the vortex to survive for thousands and up to $10^{4}$ orbits \citep{de-val-borro07a,ataiee13a,zhu14b,fu14b}. Vortices can effectively trap dust particles, leading to a more azimuthally and radially concentrated dust density distribution of larger grains at the center of the vortex \citep{Barge1995,birnstiel13a,lyra13a,ruge16a}. However, the combined effect of the dust settling and trapping inside the vortex  produces an enhanced dust-to-gas mass ratio in the vortex, leading to an increase of the dust back-reaction. This produces an alteration of the coherent vorticity pattern and destroys the vortex \citep{johansen04a,fu14a}. 
 
 Lopsided discs have been identified in high-resolution observations at (sub-)mm wavelengths. While in the case of IRS 48  \citep{vandermarel13} millimetre grains appear to be more concentrated in the horseshoe region compared to smaller sizes, in other cases (SR 21 and HD135344B; \citealt{pinilla15a}), dust trapping is not observed, further challenging the vortex scenario. HD142527 shows a large horseshoe in mm continuum emission but whether or not dust trapping occurs is more controversial \citep{perez15,muto15,casassus15a}.
 
 Here, we investigate an alternative explanation for the development of non-axisymmetric gas and dust structures based on studies of discs around black hole binaries \citep{farris14,dorazio16,ragusa16}. These showed that, for mass ratios $q\gtrsim 0.04$, the wide cavity around the primary object carved by the companion becomes eccentric and develops a strong overdensity at the cavity edge, orbiting at the local Keplerian frequency. This arises naturally even in relatively viscous discs in the presence of a sufficiently massive companion. In the protostellar case, \citet{ataiee13a} showed with 2D hydrodynamic simulations that, for lower mass ratios ($\sim10^{-3}$), the asymmetries at the cavity edge are weaker than in the vortex scenario, resulting in ring-like rather than horseshoe morphologies. It is therefore timely to explore the case with higher mass ratio to determine whether the horseshoe-like density features revealed by ALMA observations might be explained by the presence of a massive companion inside the cavity. We explore this hypothesis using global, 3D smoothed particle hydrodynamics (SPH) simulations of gas and dust evolution in a circumbinary disc, where the binary consists of a young star and either a massive planet or low mass stellar companion. We demonstrate the formation of crescent-like structures with emissivity contrast up to $\sim 10$, sufficient to explain many of the `dust horseshoes' and other non-axisymmetric features observed in transitional discs.
 
 The idea of transitional discs as circumbinary discs has been recently explored by \citet{ruiz16}. They found that the spectral energy distributions of $\sim 40\%$ of transitional discs in their sample can be explained as being produced by the flux emission of discs orbiting around binary systems.

\begin{figure*}
\begin{center}
\includegraphics[width=0.24\textwidth]{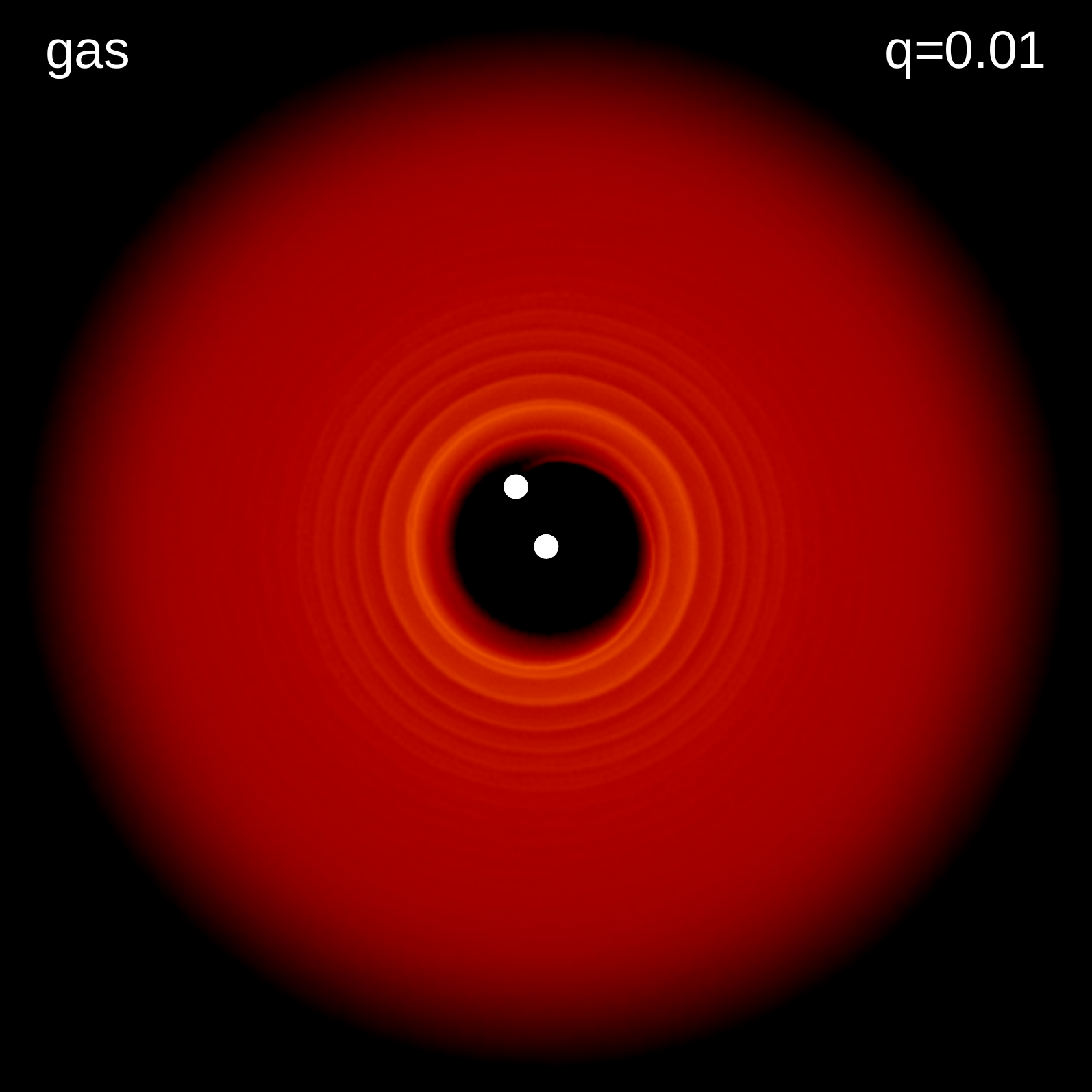}
\includegraphics[width=0.24\textwidth]{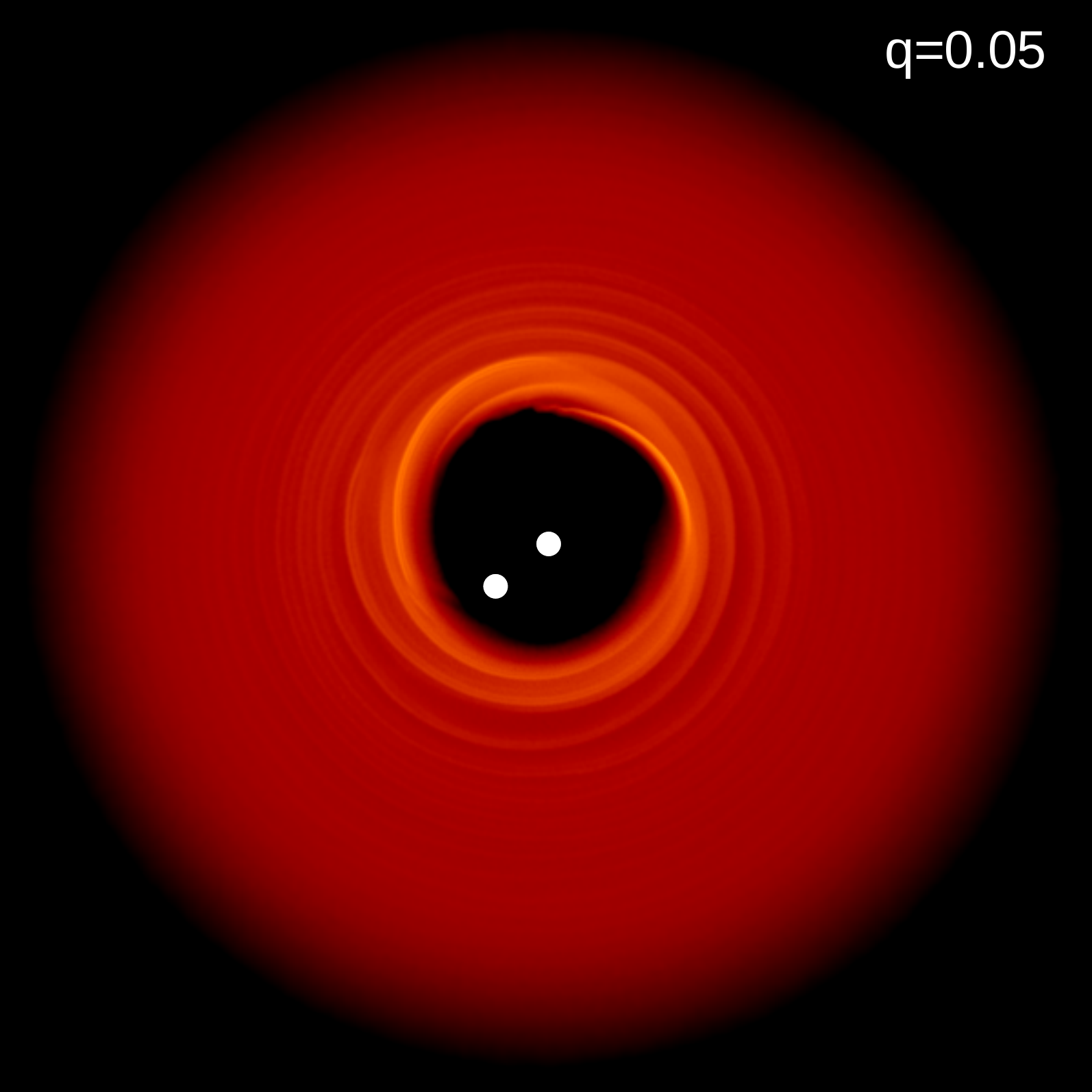}
\includegraphics[width=0.24\textwidth]{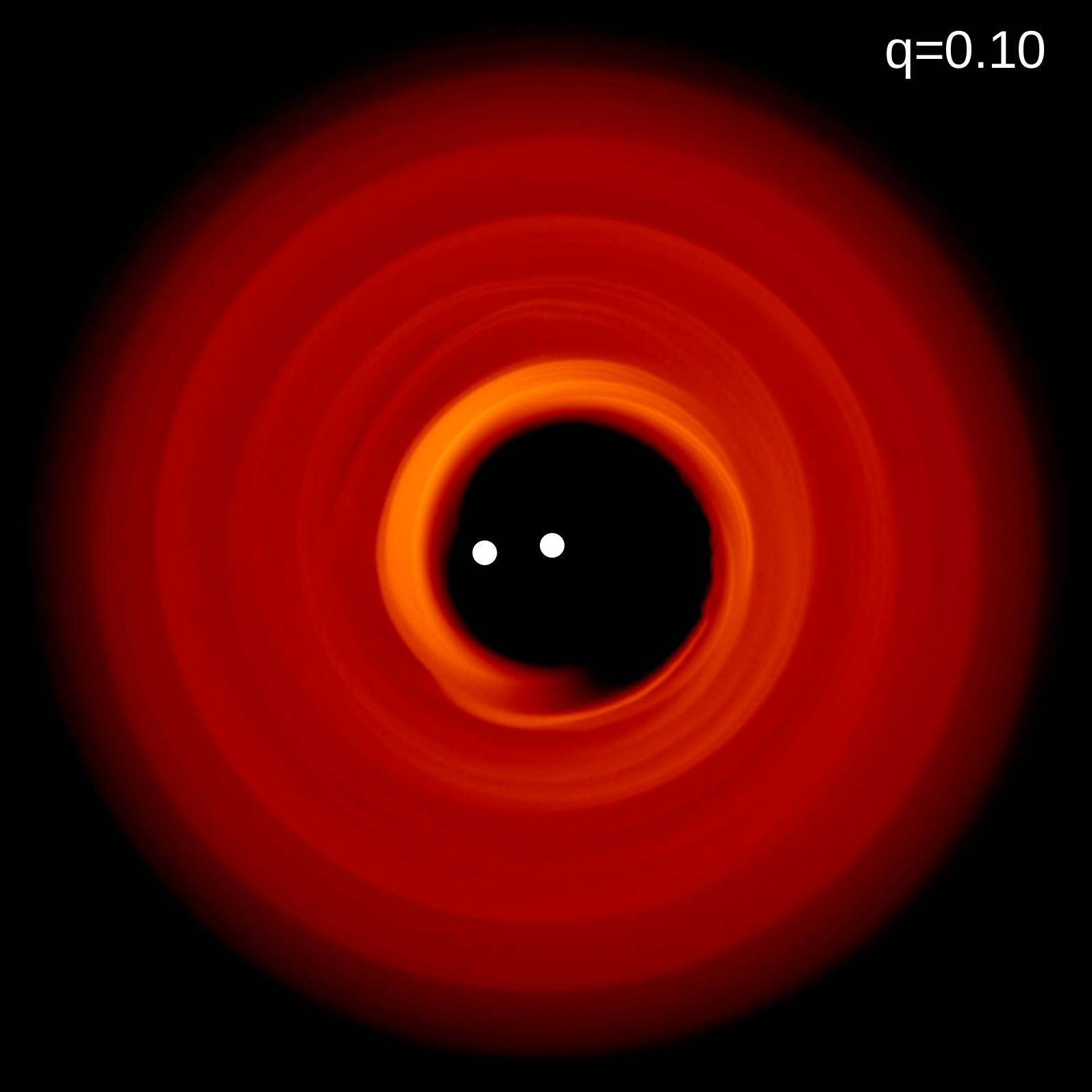}
\includegraphics[width=0.24\textwidth]{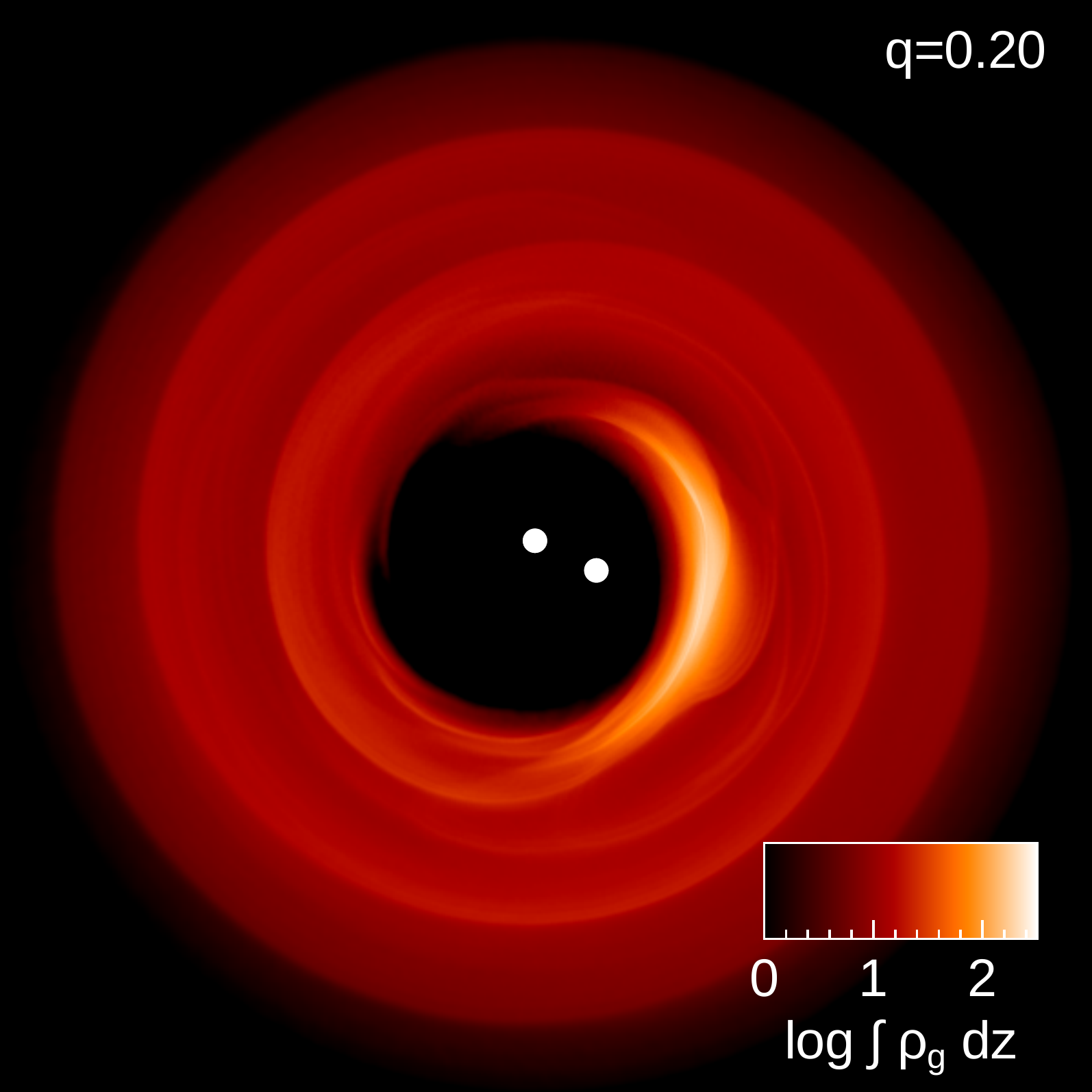}\\
\centerline{\includegraphics[width=0.24\textwidth]{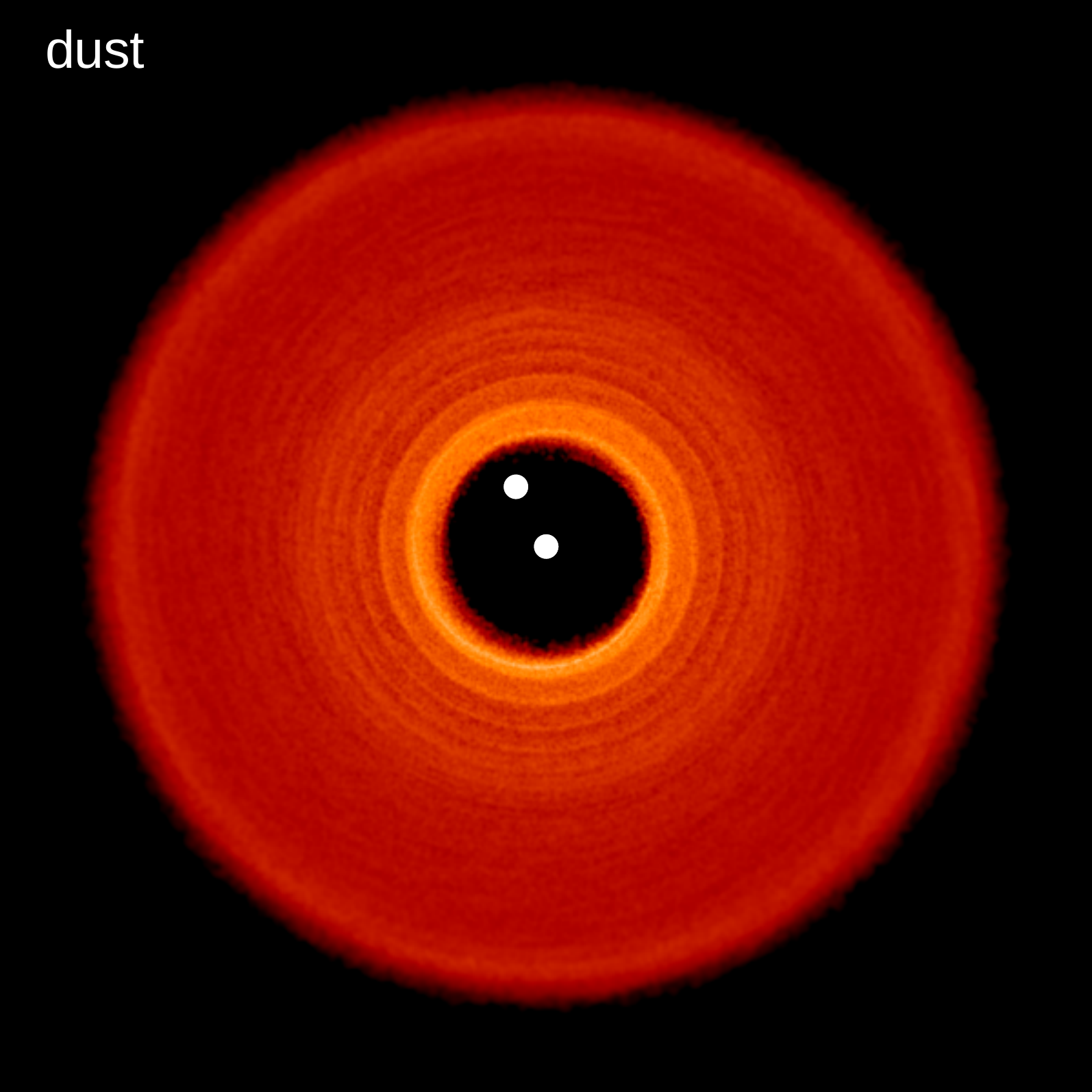}
\includegraphics[width=0.24\textwidth]{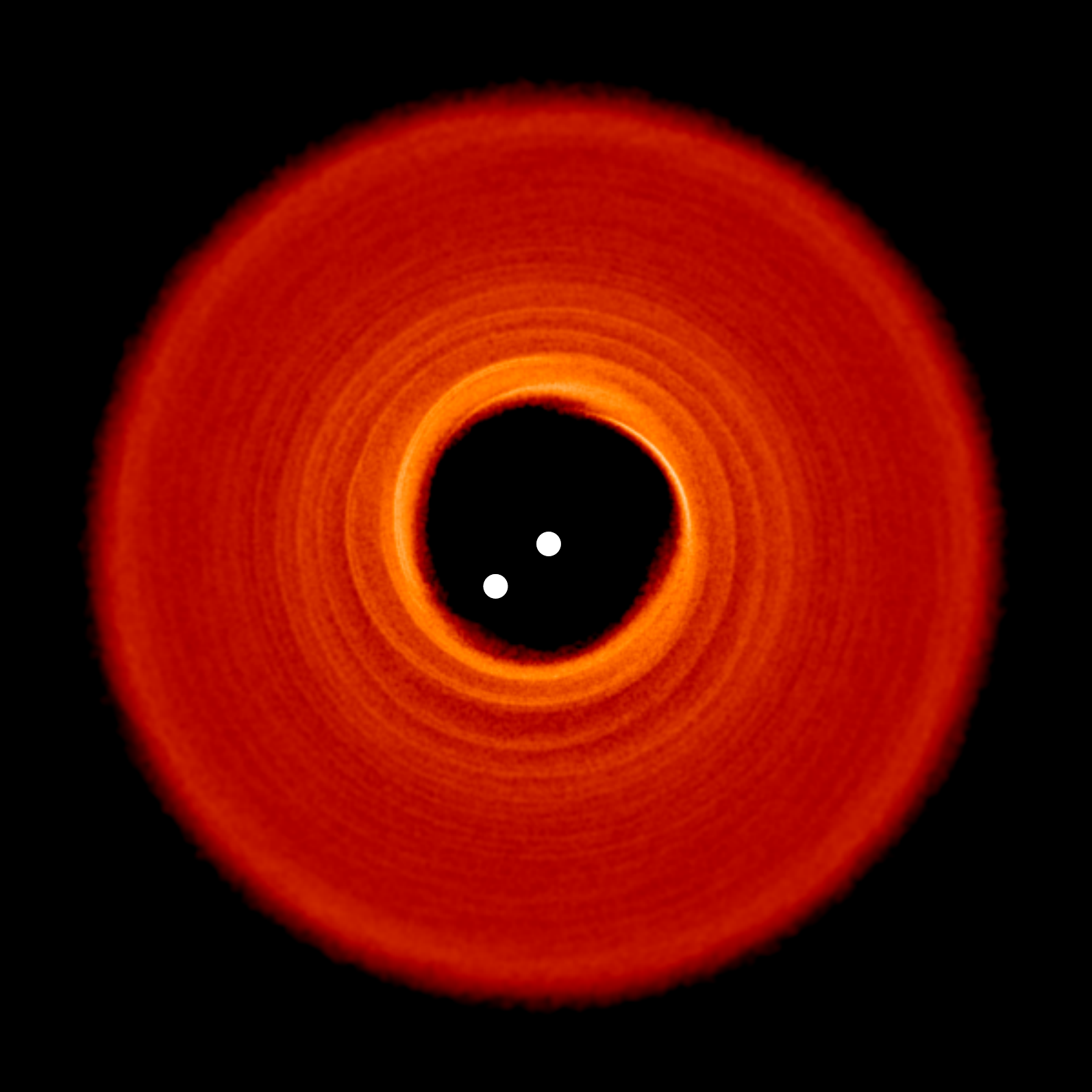}
\includegraphics[width=0.24\textwidth]{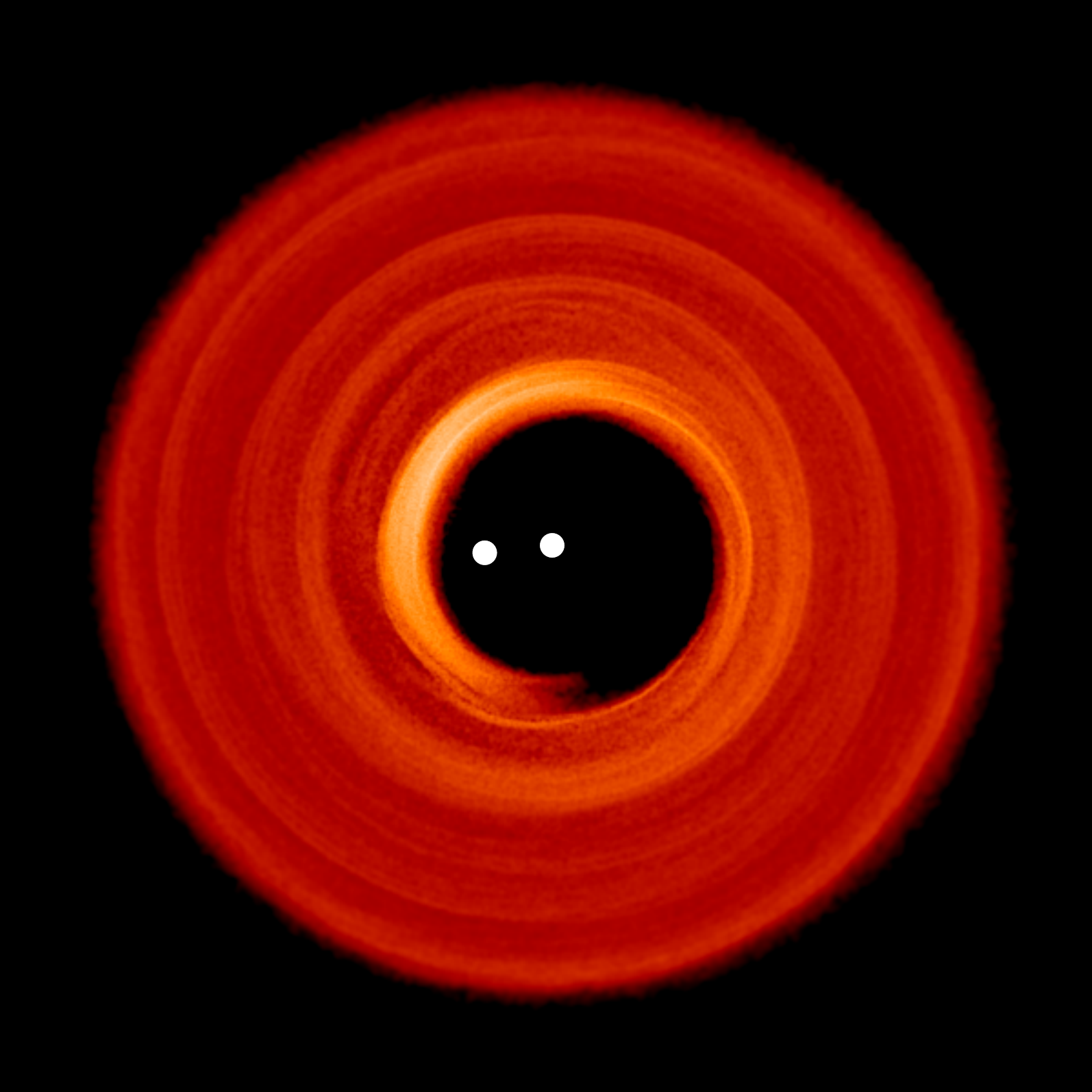}
\includegraphics[width=0.24\textwidth]{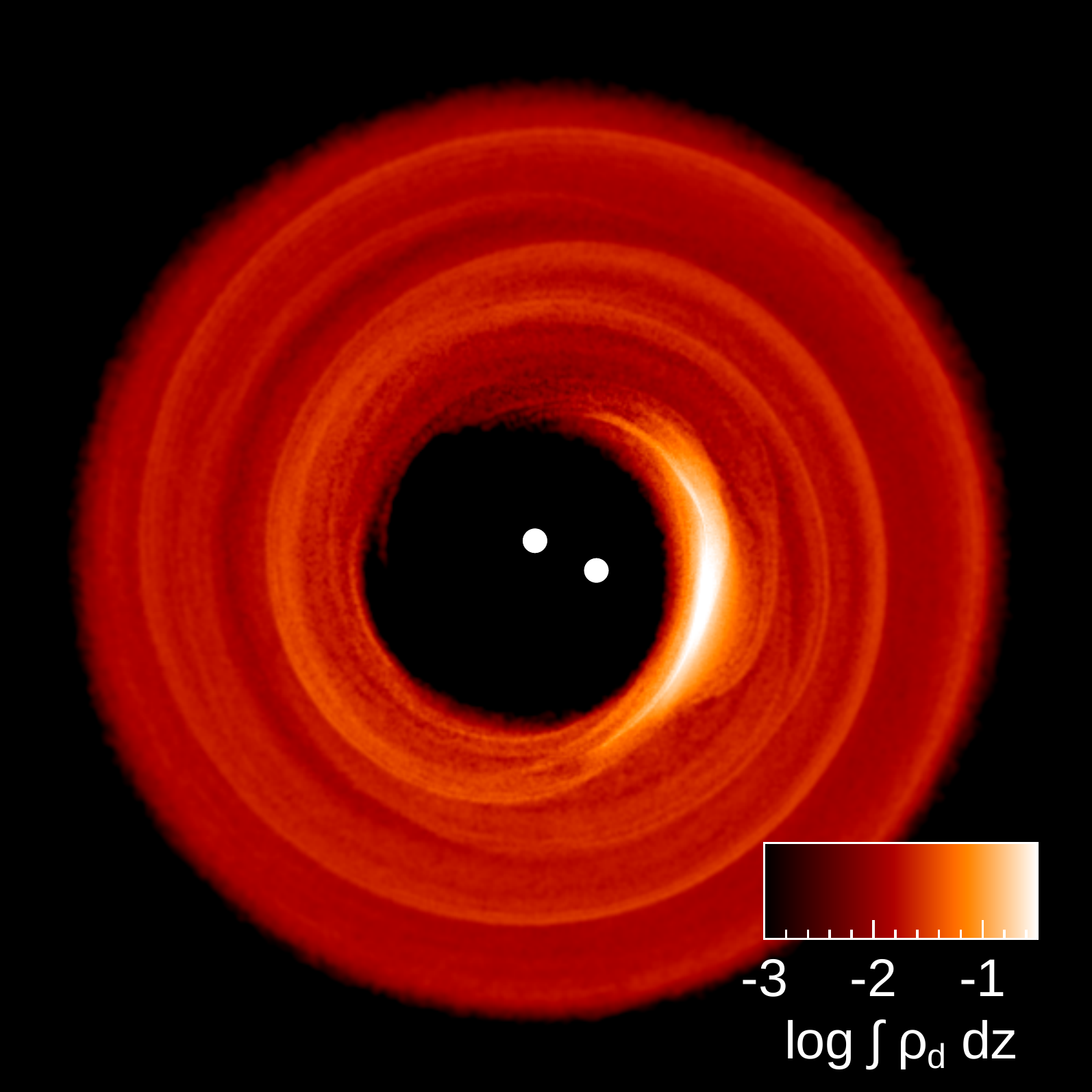}}
\centerline{\includegraphics[width=0.24\textwidth]{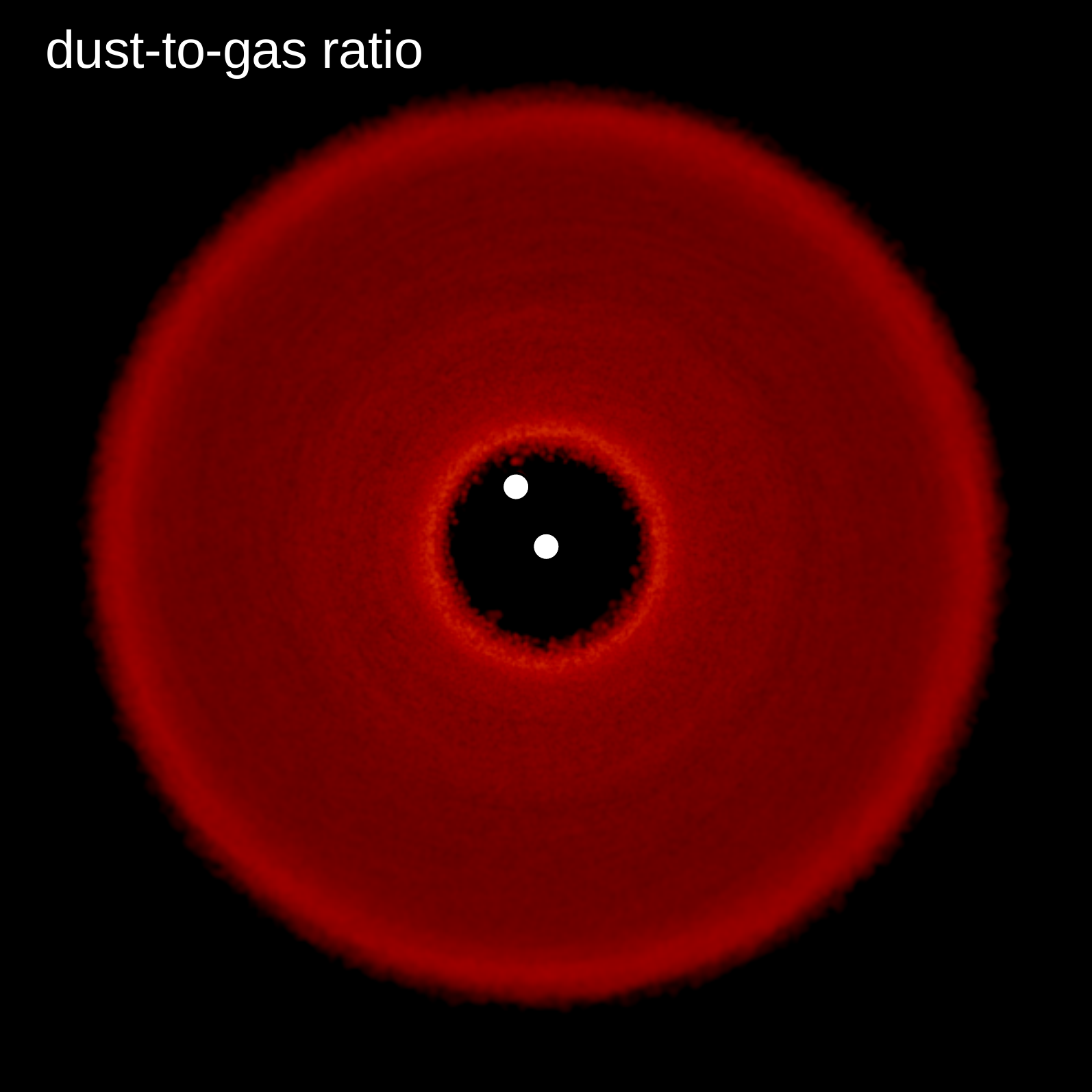}
\includegraphics[width=0.24\textwidth]{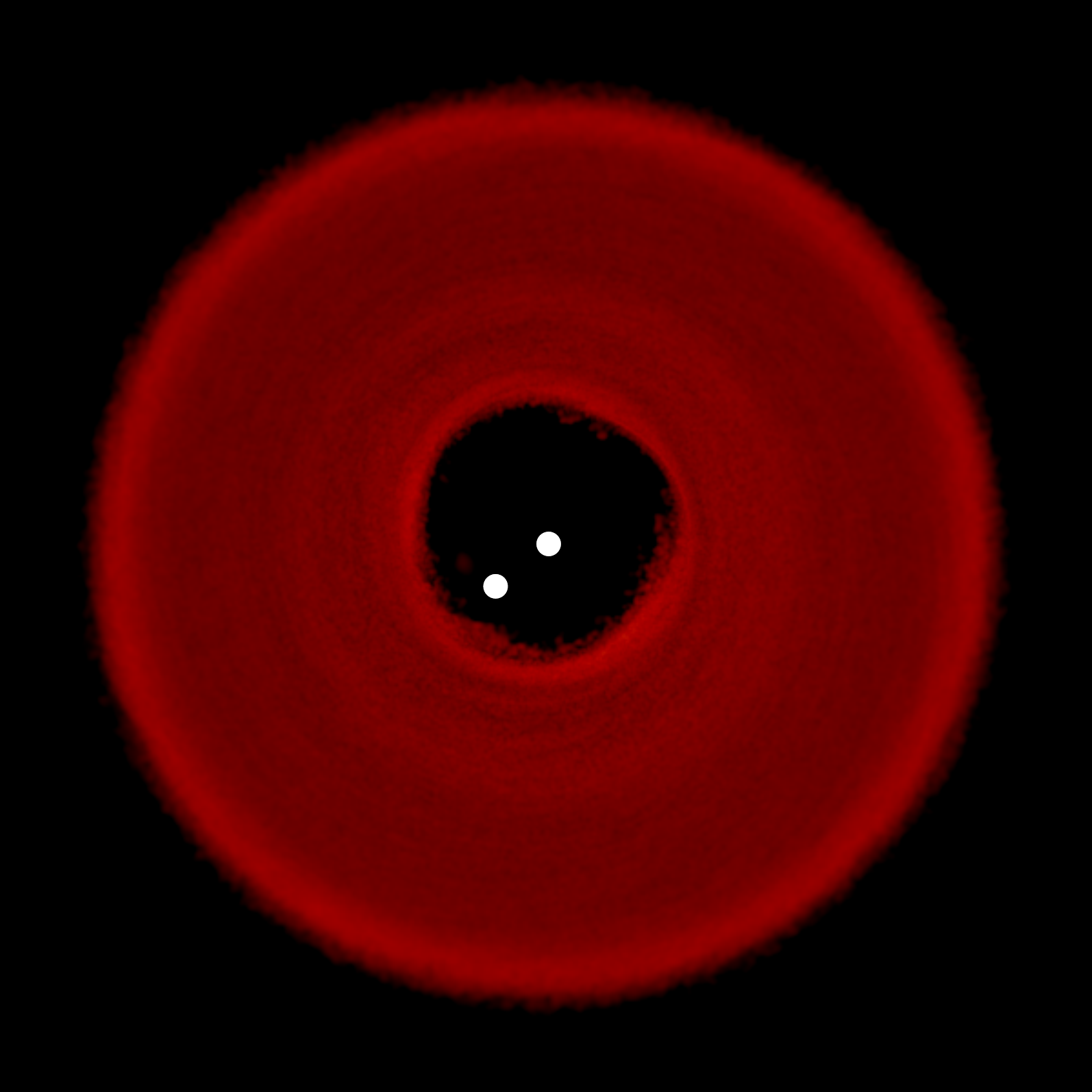}
\includegraphics[width=0.24\textwidth]{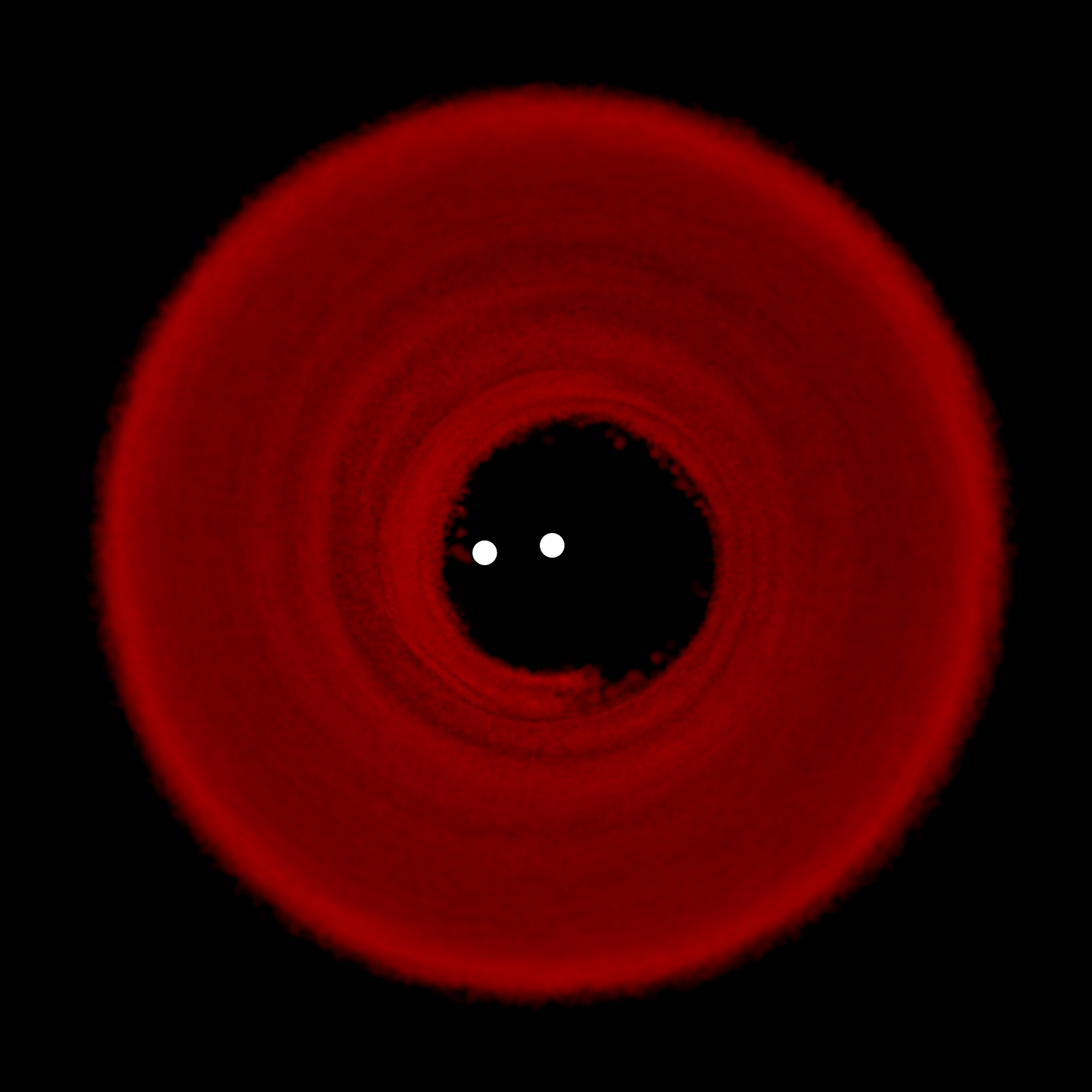}
\includegraphics[width=0.24\textwidth]{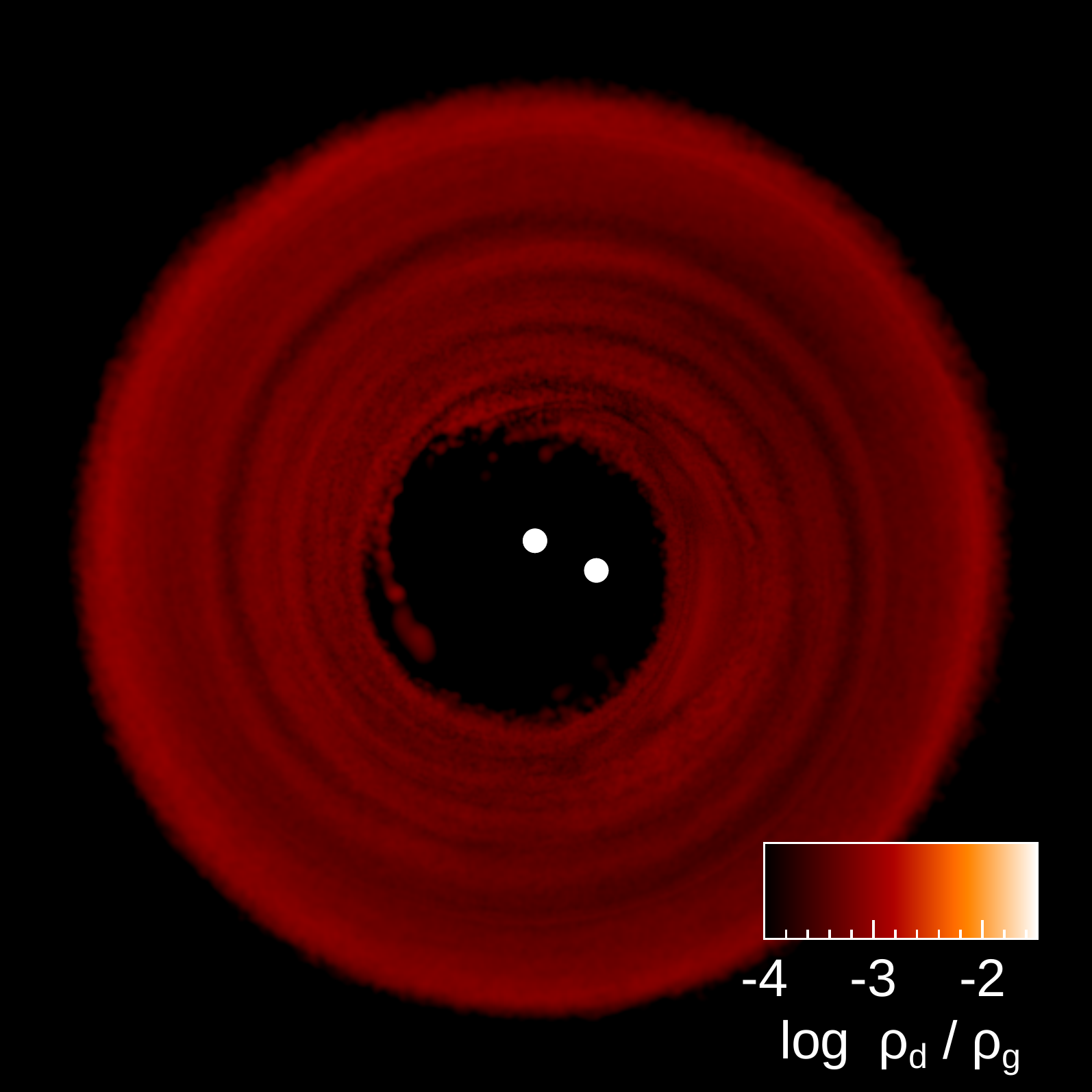}}
\caption{Gas (top row) and dust (middle row) surface density in units of g/cm$^{2}$ in logarithmic scale after 140 binary orbits for four different binary mass ratios; $q=\{0.01,0.05,0.1,0.2\}$ (left to right, respectively). High mass ratio binaries drive the formation of a large eccentric cavity leading to non-axisymmetric overdensities in both gas and dust ($q \gtrsim 0.05$; right columns). Low binary mass ratios, by contrast, produce more axisymmetric overdensities around a smaller central cavity ($q \lesssim 0.05$; left columns). The bottom row shows the column averaged dust-to-gas ratio in logarithmic scale for the different mass ratios. Note that, for the millimetre size particles we simulate, no dust trapping occurs in the overdense region. Simulated observations of these calculations are shown in Figure~\ref{fig:alma}.}
\label{fig:gasdust}
\end{center}
\end{figure*}

\begin{figure*}
\begin{center}
\includegraphics[trim={0 1cm 0 0},clip,scale=0.6]{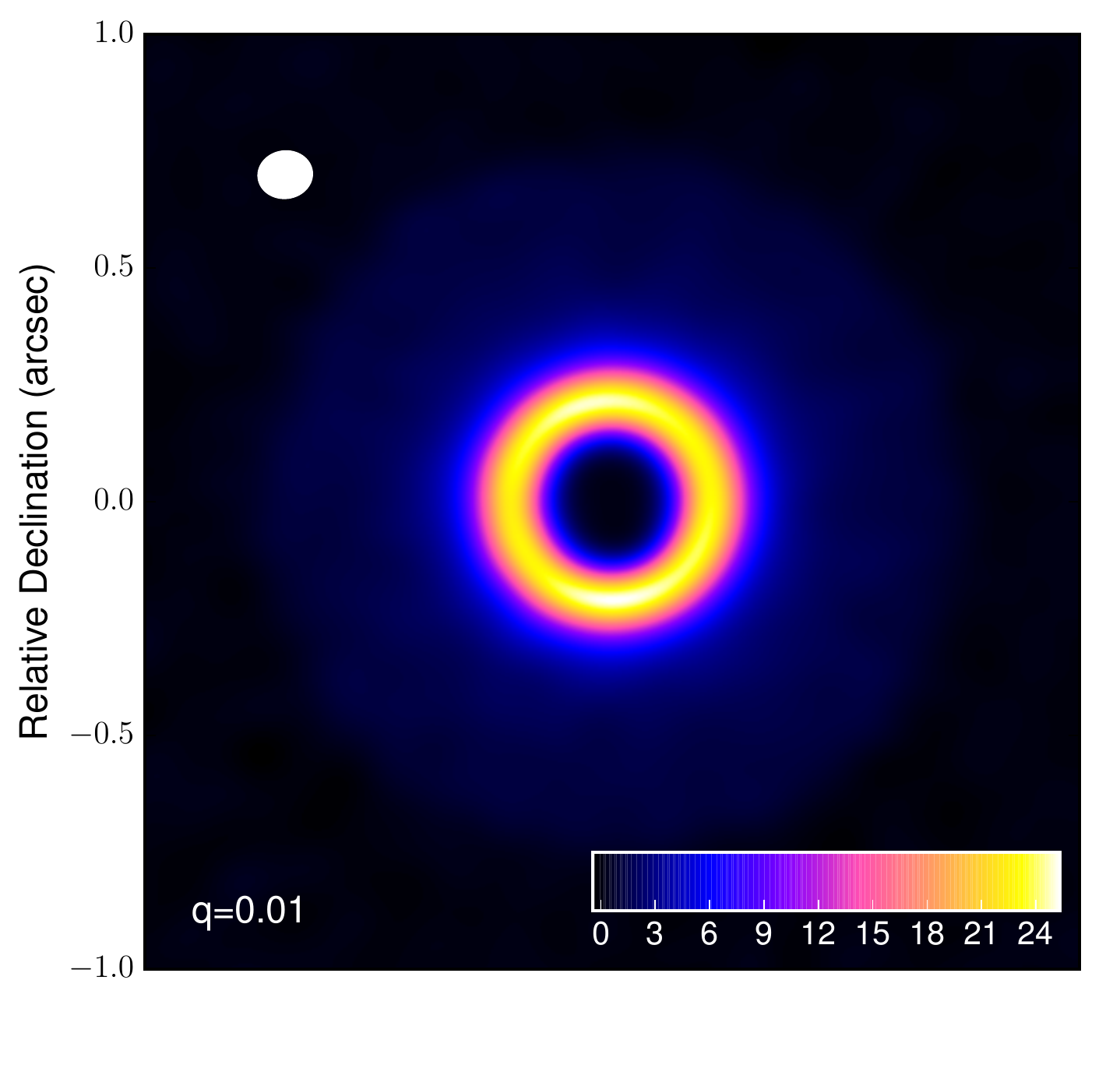}
\includegraphics[trim={1.5cm 1cm 0 0},clip,scale=0.6]{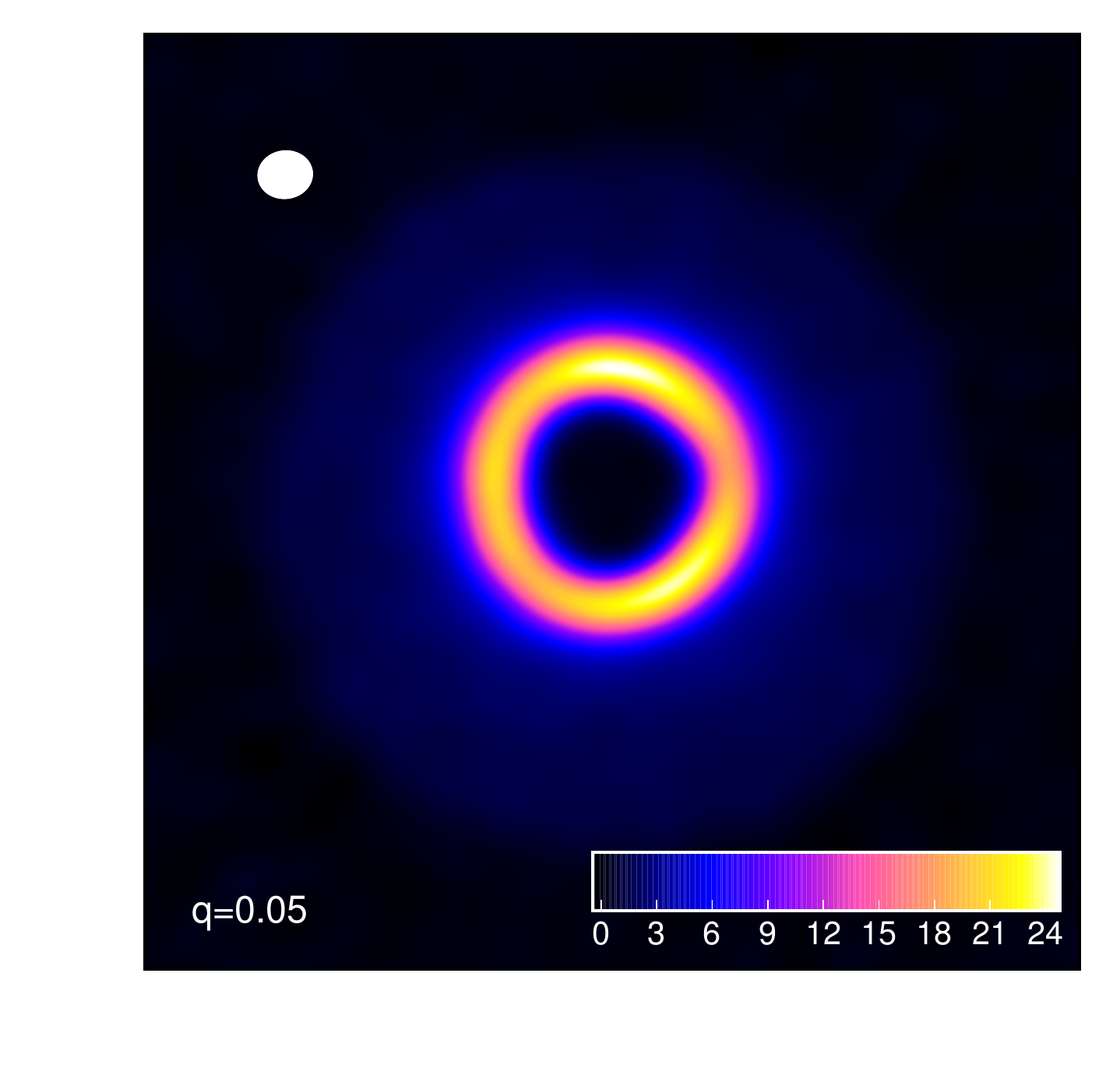}
\centerline{\includegraphics[trim={0 0 0 0},clip,scale=0.6]{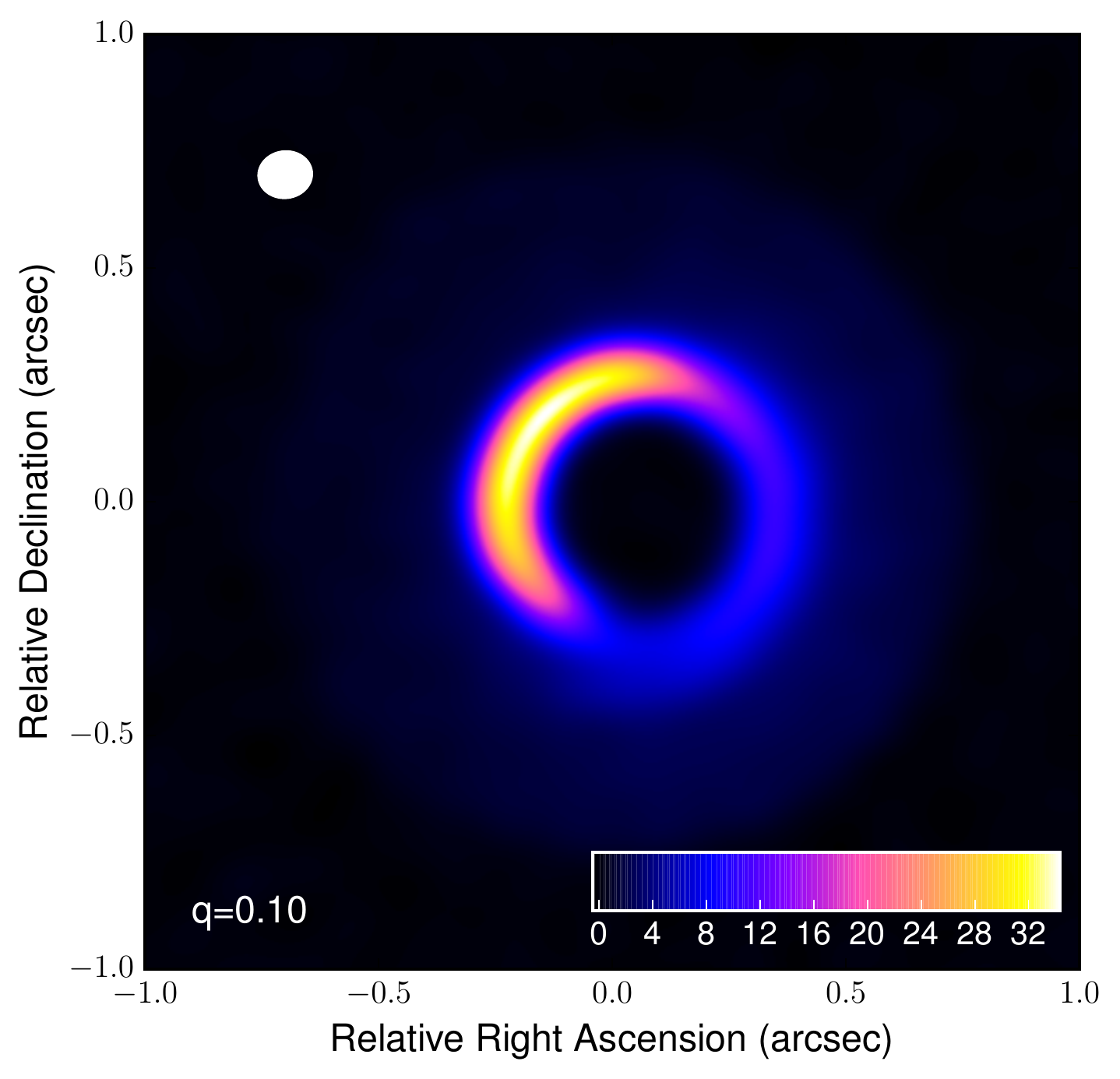}
\includegraphics[trim={1.5cm 0 0 0},clip,scale=0.6]{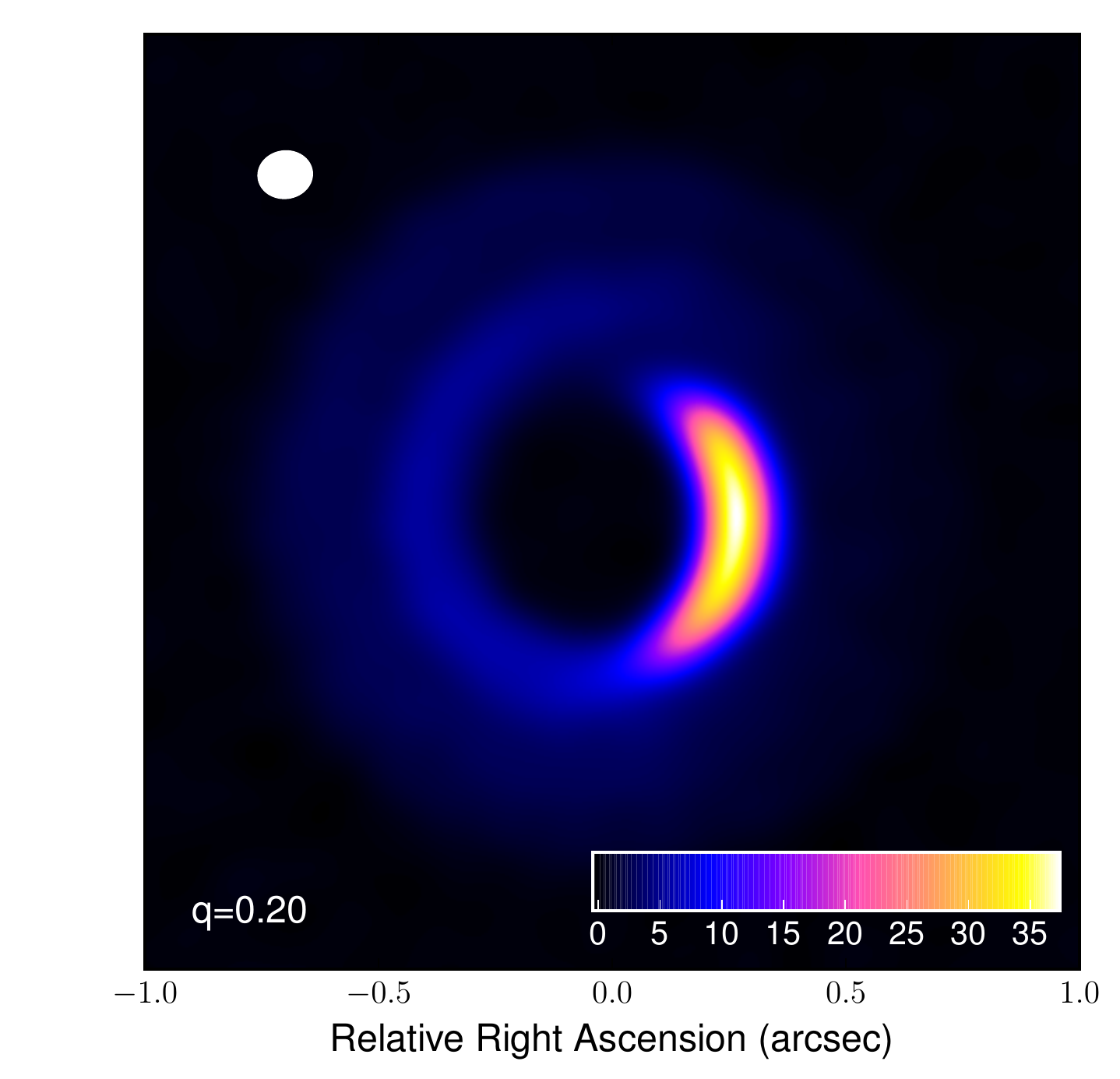}}
\caption{Comparison of ALMA simulated observations at 345 GHz of disc models with a mass ratio $q=0.01$ (upper left), $q=0.05$ (upper right), $q=0.1$ (bottom left) and $q=0.2$ (bottom right). Intensities are in mJy $\mathrm{beam^{-1}}$. The white colour in the filled ellipse in the upper left corner indicates the size of the half-power contour of the synthesized beam: $0.12\times0.1$ arcsec ($\sim 16\times 13$ au at 130 pc.).}
\label{fig:alma}
\end{center}
\end{figure*}
%
\section{Methods}
\label{sec:methods}
We perform a set of 3D gas and dust numerical simulations of a circumbinary disc surrounding a binary object (to be interpreted both as star-star or star-planet system), using the \textsc{phantom} SPH code \citep{LodatoPrice2010,PriceFederrath2010,Price2012}. The binary is represented by two sink particles \citep[e.g.][]{Bate1995,Nixon2013} that exert the gravitational force on each other and on the gas particles. The sink particles are free to move under the backreaction force of the gas on them, allowing the overall conservation of the binary-disc angular momentum, migration and eccentricity evolution of the binary. The sink particles are also allowed to accrete gas particles when they cross the sink radius and satisfy several dynamical conditions assuring that they are not able to escape from the gravitational field \citep{Bate1995}. 

We model the gas-dust interaction using the one fluid model SPH formulation developed by \citet{LaibePrice2014} and \citet{PriceLaibe2015}, assuming small grains (i.e. a Stokes number $\mathrm{St} < 1$) such that the terminal velocity approximation is valid. We set the dust grain size to be $s=1\, {\rm mm}$ and vary the mass ratio of the binary. Pressure is computed using a locally isothermal equation of state assuming a radial power-law temperature profile. 

 We exploit the SPH artificial viscosity to model the physical processes responsible for the angular momentum transfer throughout the disc. With reference to the notation used in \citet{LodatoPrice2010}, we set the artificial viscosity parameter $\alpha^{\rm AV}=0.1$ that corresponds, given our initial conditions, to a \citet{shakura73} $\alpha_{\rm SS}$ parameter ranging between $0.01\lesssim\alpha_{\rm SS}\lesssim 0.04$ across the disc. To prevent particle interpenetration we set the parameter $\beta=2$ as prescribed in \citet{Price2012}.

SPH artificial viscosity provides also a natural way to reproduce turbulent diffusion of the gas \citep{Arena2013} which is transmitted to the dust by the drag.

Each simulation is evolved for 140 binary orbits, corresponding to a physical time of $\sim 5600 \, {\rm yr}$, which is long enough to allow the dust to settle from its initial displacement and to reach quasi-stationarity in the disc shape.

\subsection{Initial conditions}
Our initial setup consists of a sink particle binary surrounded by a disc of $2\times 10^6$ SPH particles. The binary has a total mass $M_{\rm tot}=2.2 \, M_\odot$ (note, however, that the dynamics are only sensitive to the mass ratio, not the absolute mass), a binary separation $a=15\, {\rm au}$ and an orbital eccentricity $e=0$. We performed a set of four simulations varying the mass ratio $q=\{0.01;0.05;0.10;0.20\}$. 

The disc extends between an inner radius $R_{\rm in}=18\, {\rm au}$ and an outer radius $R_{\rm out}=100\, {\rm au}$, centred on the centre of mass of the binary. The  surface density distribution is $\Sigma=\Sigma_0 R^{-p}$, where $R$ is the radial coordinate in the disc, $p=0.5$ and $\Sigma_0$ determines the total disc mass, $M_{\rm disc}=(1+\epsilon) M_{\rm g,disc}$, where $\epsilon=10^{-3}$ is the millimetre dust-to-gas ratio (corresponding to a total dust-to-gas ratio of 0.01, for our assumed grain size distribution, see Section \ref{sec:alma}), and $M_{\rm g,disc}=0.05 \,M_\odot$ is the gas disc mass. Particles are distributed vertically according to a Gaussian distribution with thickness $H=c_{\rm s}/\Omega_{\rm k}$, where $c_{\rm s}$ is the gas sound speed and $\Omega_{\rm k}=\sqrt{GM_{\rm tot}/R^3}$. We assume that $H/R\propto R^{0.25}$ and that $H/R=0.05$ at $R=18$ au. The dust-to-gas ratio is $\epsilon=10^{-3}$ throughout the entire disc, implying that the dust has initially the same vertical structure as the gas. After a few orbits of the secondary, the dust has settled from its initial displacement forming a layer with thickness $H_d=H\sqrt{\alpha_{\rm SS}/\mathrm{St}}\sim 0.7\,H$, consistently with the \citet{Dubrulle1995} model.

The velocity of each particle follows a Keplerian profile centered on the binary centre of mass, with orbital velocities corrected to take account of the radial pressure gradient.

The average vertical resolution of this setup can be expressed as $\langle h/H\rangle\sim 0.2$, where $h$ is the SPH smoothing length. Since, for our parameter choice, $H_{\rm d}\sim H$, the disc remains vertically well resolved  both in the gas and in the dust.

\subsection{Simulated ALMA observations}
\label{sec:alma}
We performed mock ALMA observations of our models using the RADMC-3D Monte Carlo radiative transfer code \citep{dullemond12a} together with the Common Astronomy Software Application (CASA) ALMA simulator (version 4.5.3), focusing on ALMA band 7 (continuum emission at 345 GHz). The source of radiation is assumed to be the central star, located at the centre of the coordinate system, with $M_{\star}= 2 M_{\odot}$, $T_{\rm{eff}}=5500$ K and $R_{\star}= 2 R_{\odot}$. Dust opacities were produced using the routine\footnote{\url{https://dianaproject.wp.st-andrews.ac.uk/data-results-downloads/fortran-package}} developed by \citet{woitke16a} adopting the dust model from \citet{min16a}. Since ALMA band 7 images essentially trace millimetre particles with a maximum size of $\sim 3\lambda$ \citep{draine06a}, we assumed a dust population with a power-law grain size distribution given by $n(s) \propto s^{-m}$ between $s_{\min}=0.1 $ mm to $s_{\max}=3$ mm, with $m=3.5$. Starting from the 3D density distribution of millimetre grains of our model, we computed the spatial densities of grains in this size range by scaling the dust mass for each grain sizes according to the assumed size distribution, with a total dust mass in the size range $[0.1\,\mathrm{\mu m}, 10\, \mathrm{cm}]$  equal to 0.01 of the gas mass.

We computed full-resolution images using $10^8$ photon packages. These images were then used as input sky models to simulate realistic ALMA observations taking into account the thermal noise from the receivers and the atmosphere and assuming a perfect calibration of the visibility measurements. We assumed that all the sources were located in Ophiuchus star-forming region ($d\sim$130~pc), observed with a transit duration of 3 minutes. We assumed Cycle 2 ALMA capabilities adopting an antenna configuration that provides a beam of 0.12 $\times$ 0.1 arcsec ($\sim 16 \times 13$ au).


\section{Results}
\label{sec:results}

 Fig. \ref{fig:gasdust} shows the surface density after 140 binary orbits in the gas (top panels) and dust (middle panels) for four different disc models with increasing binary mass ratio ($q=0.01,0.05,0.1$ and $0.2$; left to right, respectively). As expected, the cavity size increases with the mass ratio \citep[e.g.][]{Artymowicz1994}. The orbital eccentricity of the gas at the cavity edge also increases with $q$, reaching $e=0.1$ for our highest mass ratio case ($q=0.2$; last right column). 
  
 For lower mass ratios ($q=0.01$ and $q=0.05$; left two columns), the dust and gas density distribution are more axisymmetric, showing a ring-like overdensity at the cavity edge. For $q\gtrsim 0.05$,  an asymmetric crescent-shaped overdensity develops at the cavity edge, with surface densities up to a factor $\sim 10$ denser than the surrounding gas, consistent with previous numerical simulations in the context of black hole binaries \citep{Dorazio2013,farris14,shi12,ragusa16}. The overdensity is a Lagrangian feature that rotates with the local orbital frequency.
 
 For fixed mass ratio, the level of contrast in the surface density across the crescent-shaped region is similar in both in the gas and in the dust. This is due to the fact that the high gas density in the lump produces a strong aerodynamical coupling between the gas and the dust in the disc. 
Interestingly, the sharpness of the region increases with increasing mass ratio. 

 Fig. \ref{fig:alma} shows mock ALMA images of our disc models at band 7 for the four different mass ratios. The simulated ALMA images reflect the density structures observed in Fig. \ref{fig:gasdust}. In particular, a crescent or `dust horseshoe' is evident for $q> 0.05$, with the contrast increasing with increasing mass ratio: for $q=0.1$ the typical contrast is $\approx 5$, while for $q=0.2$ we obtain a contrast $\approx 7$. For $q=0.05$, the ALMA image shows a double-lobed feature with a low contrast $\sim 1.5$, similar to those observed in SR21 or DoAr 44 \citep{vandermarel16a}. For $q=0.01$ a ring-like structure can be observed, as observed e.g. in Sz 91 \citep{canovas16}.

\begin{figure}
\begin{center}
\includegraphics[width=0.49\columnwidth]{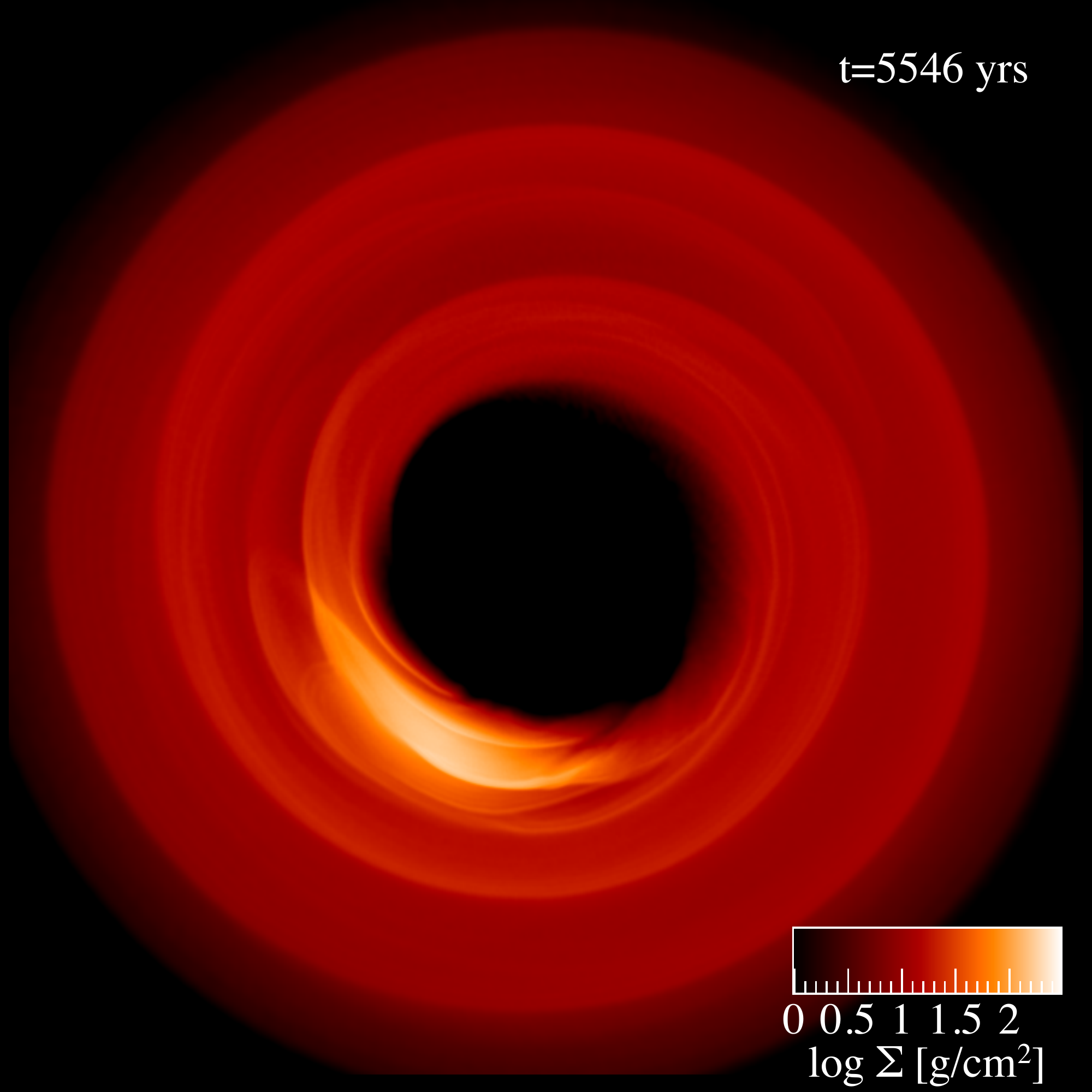}
\includegraphics[width=0.49\columnwidth]{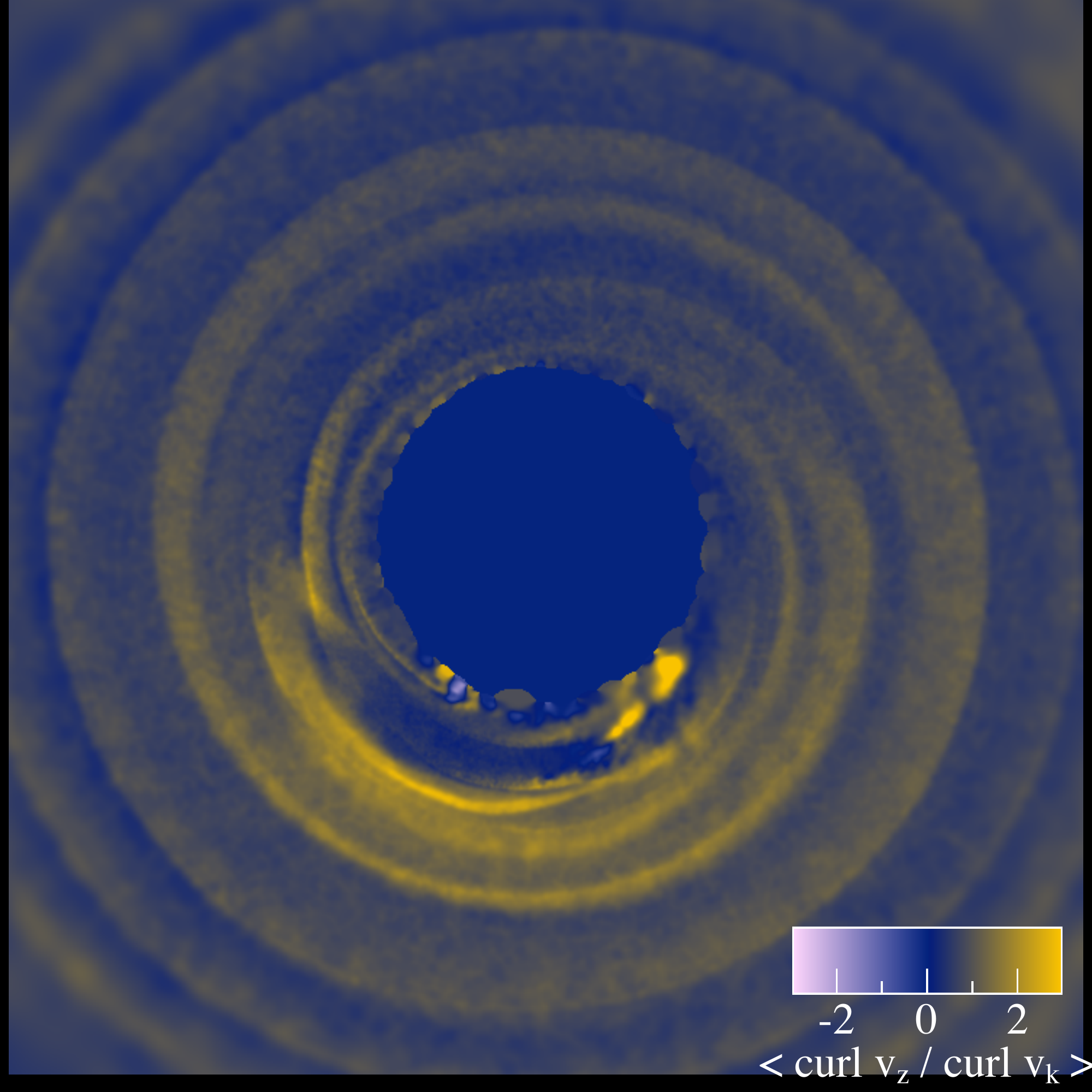}
\caption{Map of the vorticity ${\bm \omega}=\nabla\times {\bf v}$ (right panel) in the region of the overdensity, scaled to the Keplerian value ${\bm \omega}_{\rm K}=\nabla\times {\bf v}_{\rm K}$, for $q=0.2$. The left panel shows the gas density structure. There is no evidence of vortices associated with the overdense region (cf. \citealt{owen16a}).}
\label{fig:velocity}
\end{center}
\end{figure}

The right panel of Fig. \ref{fig:velocity} shows a snapshot of the vorticity ${\bm \omega}=\nabla\times {\bf v}$, scaled to the Keplerian value ${\bm \omega}_{\rm K}=\nabla\times {\bf v}_{\rm K}$, where $v_{\rm K}$ is the keplerian velocity field. The flow is close to keplerian in the outer regions of the disc, while in the overdense region the value of the vorticity is $0\lesssim {\bm \omega/\bm \omega}_{\rm K}<1$. The extended region outside the overdense crescent where ${\bm \omega}>{\bm \omega}_{\rm K}$ is due to the steeper than keplerian gradient of the azimuthal velocity. Vortices induced by the Rossby wave instability typically result in much higher vorticities, with anti-cyclonic vortices reaching $| {\bm \omega/ \bm \omega}_{\rm K}|\sim 2$ \citep{owen16a}.

\section{Discussion}
\label{sec:discussion}

The idea that large scale asymmetries might be due to a planetary companion was explored by \citet{ataiee13a}, who concluded that planetary mass objects only produce ring-like features in the disc, in contrast to the observed horseshoe. However, we have shown the dynamics induced in the disc by low and high mass companions is markedly different. It is known that low-mass companions, with $q\sim 10^{-3}$ can produce eccentric cavities, that precess slowly around the star-planet system \citep{papaloizou01,kley06}. In contrast, more massive companions, with $q\gtrsim 0.04$ \citep{shi12,dorazio16} produce strong non-axisymmetric lumps that orbit at the local Keplerian frequency. We have explored the latter case in this paper. For sufficiently massive companions (binary mass ratio $q=0.2$) we obtain an azimuthal contrast of the order of $\sim 10$ in mm-wave map, with the contrast an increasing function of the binary mass ratio.

The mechanism causing the formation of the gas overdensity at the cavity edge is still unclear \citep{shi12}, but is thought to be related to shocks in the gas at the cavity edge, arising from the intersection of gas flows within the cavity.  Thus, it might be expected that the chemistry would be affected by shocks. Processes such as desorption of various chemical species from the surface of disc dust grains and gas-phase chemical reactions due to shocks occurring in the cavity wall, produce clear chemical signatures of the disc dynamics (see e.g.~\citealt{ilee11a} in the case of shocks induced by gravitational instabilities) which may be detected using ALMA. Additionally, shocks might induce the emission of forbidden lines, the detection of which would confirm eccentric cavities as the origin of these structures.

As previously mentioned, horseshoes in transitional discs are often assumed to be due to a vortex induced by a low-mass companion in the cavity. In this case, a variation of the azimuthal extenst of the horseshoe at different wavelengths is expected. Indeed, models of dust trapping by a vortex predict that larger grains would be more azimuthally concentrated in the centre of the vortex. However, in some cases (SR21 and HD135344B \citealt{pinilla15a}) smaller grains appear to be more trapped than larger grains. Since we compute the dynamics of a single species of dust, we cannot predict if our model would reproduce this scenario.

\subsection{Comparison with observed systems}

Non-axisymmetric features have been observed in a handful of transitional discs (see Table \ref{tab:list}).
We report here below a brief summary about the cavity features and the current evidence for the presence of massive companions in these systems. 

For what concern the upper-limits on the mass of putative companions, the most accurate results have been obtained applying the aperture masking interferometric observations and speckle imaging in the near-IR waveband. 
It is worth noticing that detecting planets through imaging is challenging due to the proximity of planets to the central star and their low contrast ratio in emission compared to the brightness of their host star. Additionally, massive companions might have eccentric or misaligned orbits with respect to the disc. This implies that the size of the cavity they are able to carve can be much larger than $\sim 2$ times the separation at which they are located in the imaging due to projection effects or orbital phase (a planet might not be resolved at the pericentre of its orbit, while resolved and thus detectable at the apocentre). 

\subsubsection{HD135344B}

Recent observations of line and continuum emission from HD135344B evidenced the presence of a cavity both in the gas ($\sim 30$ au) and in the dust ($\sim 40$ au) \citep{vandermarel16a}. The continuum emission shows also a well defined crescent shaped overdense feature at the cavity edge with a mild contrast \citep{vandermarel16a}.
A spiral structure has also been detected in the near-IR scattered light, constituting a strong indication of the presence of a massive companion \citep{garufi13}. Using the ``locally optimized combination of images'' (LOCI) technique, in order to be able to possibly resolve and locate the exact position of the companion, \citet{vicente2011} put an upper-limit of $M_c\sim 230 M_{\rm J}$ at separations $a \lesssim 14$ au and $M_c\sim 85 \,M_{\rm J}$ at $a\lesssim 37$ au.

Given the central star estimated mass $M_\star\sim 1.7 M_\odot$, the upper-limits on the secondary mass imply mass ratios $q\lesssim 0.05$ at $a \lesssim 37$ au and up to $q\sim 0.13$ for separations $a \lesssim 14$ au. This is consistent with our models, since the crescent-like feature with contrast $\lesssim 10$ observed in HD135344B \citep{vandermarel16b} is similar to what we obtain for our $q=0.1$ case (bottom left panel of Fig.~\ref{fig:alma}).

\subsubsection{SR 21}

The continuum emission from this system shows different asymmetric features at different wavelengths: a crescent shaped overdense feature at $690$ GHz \citep{perez14}, and a double-lobed structure at 345 GHz \citep{vandermarel16a}. In both cases the contrast is mild ($\lesssim 10$). Modeling the dust emission \citet{vandermarel16a} inferred a cavity edge in the dust at $\sim 25$ au, while the gas cavity appears to be much smaller ($\sim 7$ au, \citealp{pontoppidan2008}).

The presence of a warm companion surrounded by a cloud of accreting gas in this system was invoked by \citet{eisner2009} to explain an excess in the near-IR and mid-IR SEDs, which could be explained by an additional warm ($\sim 700\, {\rm K}$) black body emission from an extended region of $~40 \,R_\odot$. The total luminosity produced by the companion in this framework appears to be consistent with a T-Tauri star with mass $M_c\sim 0.2\,M_\odot$ enveloped in a gaseous cloud \citep{follette2013}. Using the angular differential imaging technique to increase the image resolution, \citet{follette2013} was able to rule out the presence of secondary stellar object for separations $a \gtrsim 18$ au; based on the contrast sensitivity achieved by \citet{follette2013}, \citet{wright15a} constrained the upper-limit on the companion mass to $\sim 40-60\, M_{\rm J}$ at separations $a \gtrsim 18$ au, implying that such a stellar source needs to be located at separations $a \lesssim 18 $ au. This implies possible secondary-to-primary mass ratios of $q\sim 0.1$ for $a \lesssim 18$ au and $q\sim 0.03$ for $a \gtrsim 18$ au. 
The double-lobed structures detected in the continuum emission in SR 21 \citep[e.g.][]{pinilla15a} at 345 GHz are consistent with our models with $q\sim0.05$, which is in agreement with the detection limits reported in literature.

\subsubsection{IRS 48}
With an azimuthal contrast of $\gtrsim$ 130 of the peak emission compared to the background disc, IRS 48 represents the source with the strongest crescent-shaped dust structure in our sample. This dramatic azimuthal range at the cavity wall observed in IRS 48 is best described by the segregation of millimetre grain sizes induced by an azimuthal bump in pressure \citep{vandermarel13}. The dust cavity was found to be extended 60 au from the central star \citep{bruderer14a}.
The continuum asymmetry has been modeled as a major dust trap, triggered by the presence of a substellar companion with a mass of 9$M_{\mathrm{J}}$ \citep{zhu14b}. This would seem very unlikely to correspond to an equally sharp gas distribution.
Alternatively, \citet{wright15a}, based on the detection limit reported in \citet{ratzka05a}, rule out a potential companion with a mass $\gtrsim 100\,M_{\mathrm{J}}$, which correspond to a mass ratio of $q\sim0.05$ at a radial separation of 19 au. As expected, this result does not appear consistent with the mechanism presented in this paper.

\subsubsection{DoAr 44}
 DoAr 44 (also known as ROX 44 and Haro 1-16) is the source with the mildest contrast in (sub-) mm continuum emission in our sample \citep{vandermarel16a}. This source has been classified as a pre-transitional disc with a dust cavity between 2 and 32 au \citep{espaillat10a,vandermarel16a}. Based on the companion detection limits reported in 
\citet{ratzka05a}, \citet{wright15a} derived that a potential companion should have a mass $\lesssim 80 \,M_{\mathrm{J}}$ at a separation $\lesssim 12.5$ au which correspond to a mass ratio of  $\lesssim 0.06$, adopting the star properties reported in \citet{espaillat10a}. The values of the mass ratio inferred from observations appear to be consistent with our scenario:  the double-lobed structures observed in DoAr 44 can be explained for $q\sim 0.05$ according to our model.

\subsubsection{HD142527}

HD142527 harbours a disc with a wide dust cavity extending from 10 au to 120 au. At the cavity inner wall,  the (sub-) millimetre dust continuum emission show a bright horseshoe with contrast $\sim$30.
The existence of a massive close companion with $q\approx 0.16$ has been established for HD142527 using the sparse masking aperture technique \citep{biller12,lacour16a}. This is particularly interesting since the contrast in HD142527 of $\sim 30$ is within a factor of 3 of the contrast we find in our highest mass ratio.
However, this companion is also inclined by $\sim 70^{\circ}$ with respect to the disc \citep{lacour16} and this case might be further complicated by a strong warp \citep{casassus15}.  Intriguingly, \citet{casassus15a} comment that ``\emph{the large sub-mm crescent [in HD142527] mostly reflects the gas background, with relatively inefficient trapping, so that the observed contrast ratio of $\sim$ 30 is accounted for with a contrast of 20 in the gas}'', consistent with our model. Seemingly this conflicts with \citet{muto15} who found variations of $\sim 10$--$30$ in the dust-to-gas ratio. This difference may be explained by uncertainties in grain surface chemistry, in particular whether or not a fraction of CO is depleted on dust grains \citep{casassus16a}.

\subsubsection{Lk H$\alpha$ 330}

Lk H$\alpha$ 330 is characterised by a millimetre dust cavity with a size of about 40 au and an azimuthal intensity variation of a factor of two. Recent observations performed by \citet{willson2016}, using the sparse aperture masking technique in the K' near infrared band, on Lk H$\alpha$ 330 revealed the presence of a possible massive companion characterized by a value of $M_{\rm c}\dot M_{\rm c}\sim 10^{-3}\, {\rm M_{\rm J}^2yr^{-1}}$ orbiting at a separation $a\sim 37\, {\rm au}$ from the central star. Assuming an accretion rate on the secondary object of $\dot M_{\rm c}\lesssim 10^{-8} \, M_\odot {\rm yr^{-1}}$ implies an mass $M_{\rm c}\gtrsim 100 \,{\rm M_{\rm J}}$ and a mass ratio $q\gtrsim 0.05$. It should be noted that previous works \citep{brown2009,andrews2011,isella2013} had reported private communications that, based on near-IR observations, ruled out the presence of secondary objects with masses $M_c\gtrsim 50\, {\rm M_j}$ ($q\gtrsim 0.025$) at separations $M_c\gtrsim 10\, {\rm au}$, indicating how elusive these objects might be. In any event, these estimates for the companion mass appear to be consistent with the low sharpness of the crescent-shaped structure predicted by our model (see the upper panels of Fig.~\ref{fig:alma}).

\begin{table}
\begin{center}
\begin{tabular}{|c|c|c|c|c}
Name & Contrast & Dust trapping & Companion & Consistency \\
\hline
HD135344B & $\lesssim 10$ & No & Strong indication & Yes\\
SR 21 & $ \lesssim 10$ & No & Indication & Yes\\
DoAr 44 & $ \lesssim 10$ & ? & ? & Yes\\
IRS 48 & $ \gtrsim 100$ & Yes & ? & No\\
HD142527 & $\sim 30$ & $\textrm{cm grains?}$ & Yes & Yes\\
Lk H$\alpha$ 330 & $ \lesssim 10$ & ? & Indication & Yes\\
\hline
\end{tabular}
\caption{Summary of transition discs displaying horseshoe or other non-axisymmetric features. For each source, we indicate the observed contrast in mm images, whether there is evidence for dust trapping in the crescent, and whether the system is known to host a massive companion. The last column indicates whether the observed structures are consistent with our model, given the upper-limits on the companion mass as reported in the literature.}
\label{tab:list}
\end{center}
\end{table}%

\section{Conclusions}
\label{sec:conclusions}

 We performed 3D SPH gas and dust simulations of circumbinary discs surrounding a protostar and either a low mass stellar companion or massive protoplanet. We showed that the companion carves a wide, eccentric cavity in the disc, resulting in a non-axisymmetric gas overdensity at the cavity edge. For sufficiently large binary mass ratios this feature appears as a `horseshoe' in millimetre wavelength dust continuum images, as observed in several transition discs. 

 Our model makes testable predictions that can be used to observationally distinguish the eccentric cavity model from the more commonly assumed `gap edge vortex' model. We identify the following main features differentiating the two processes, which can be used as the basis for observational tests of our hypothesis:
\begin{enumerate}
\item Dust and gas kinematics. In our model, the fluid velocity is close to Keplerian, and does not show the large vorticity expected in the vortex model (see Fig. \ref{fig:velocity}).
\item Our mechanism applies both in high and in low viscosity discs, while vortices only arise for $\alpha\lesssim 10^{-4}$. 
\item Dust horseshoes arise from eccentric cavities only for relatively large mass ratios $q\gtrsim 0.04$, while in principle vortices can arise for lower mass planets. Clearly, establishing whether a relatively massive companion is present within the cavity is a key observational test of our model. 
\item The structures described in this paper only occur at the edge of the central cavity, while vortices can occur in principle at any location within the disc.
\item Less massive companions should produce more axisymmetric structures, potentially explaining the `dust rings' seen in (e.g.) Sz 91 and DoAr 44. We predict that a higher degree of non-axisymmetry around larger central cavities.
\item Our model does not require azimuthal dust trapping and the observed contrast largely reflects the gas density contrast. 
\end{enumerate}
In summary, cavities opened by massive companions are a promising mechanism for explaining rings, lopsided features and horseshoes seen in transition discs.

\section*{Acknowledgments}
We acknowledge the anonymous referee for constructive comments and useful suggestions that improved the paper. ER, GD and G. Lodato acknowledge funding via PRIN MIUR prot. 2010LY5N2T. G. Laibe is funded by ERC FP7 grant ECOGAL. DJP acknowledges funding via an Australian Research Council Future Fellowship FT130100034 and Discovery Project DP130102078. We used SPLASH \citep{price07a}. We thank S. Casassus, A. Sicilia-Aguilar, N. Van der Marel and J. Cuadra for useful discussions.

\label{lastpage}

\bibliography{biblio}

\end{document}